\documentclass[aps,prl,twocolumn,superscriptaddress,amsmath,amssymb]{revtex4}
\usepackage{dcolumn}
\usepackage{bm}
\usepackage{graphicx}
\usepackage{subfigure}
\usepackage{epstopdf}
\usepackage[makeroom]{cancel}
\usepackage[usenames,dvipsnames,svgnames]{xcolor}

\begin{document}
\title{Probing chiral edge states in topological superconductors through spin-polarized local density of state measurements}
\author{Kristofer Bj\"{o}rnson}
\affiliation{Department of Physics and Astronomy, Uppsala University, Box 516, S-751 20 Uppsala, Sweden}
\affiliation{Niels Bohr Institute, University of Copenhagen, Juliane Maries Vej 30, DK-2100 Copenhagen, Denmark}
\author{Annica M. Black-Schaffer}
\affiliation{Department of Physics and Astronomy, Uppsala University, Box 516, S-751 20 Uppsala, Sweden}

\begin{abstract}
We show that spin-polarized local density of states (LDOS) measurements can uniquely determine the chiral nature of topologically protected edge states surrounding a ferromagnetic island embedded in a conventional superconductor with spin-orbit coupling. The spin-polarized LDOS show a strong spin-polarization directly tied to the normal direction of the edge, with opposite polarizations on opposite sides of the island, and with a distinct oscillatory pattern in energy.
\end{abstract}

\maketitle

The past few years have seen a rapid development of the field of topological superconductivity \cite{PhysUsp.44.131, PhysRevLett.100.096407, NatPhys.5.614, RevModPhys.82.3045, RevModPhys.83.1057}.
Motivated both by exploration of a new frontier in physics and the prospect of using Majorana bound states for topological quantum computation \cite{PhysUsp.44.131, RevModPhys.80.1083, RepProgPhys.75.076501}, substantial theoretical and experimental progress have occurred \cite{Science.336.1003, NatPhys.8.795, NatPhys.8.887, Science.346.602, npjQuantumInformation.2.16035, PhysRevLett.115.197204, RepProgPhys.80.076501}.
While many topological superconductors are predicted to host Majorana bound states, they only arise under specific geometric conditions, such as when a topological superconductor forms a one-dimensional (1D) wire or in the superconducting vortex core of a 2D system \cite{PhysRevLett.103.020401, PhysRevLett.104.040502, PhysRevLett.105.077001, PhysRevLett.105.177002, PhysRevB.82.134521, PhysRevB.88.024501, PhysRevB.91.214514, PhysRevB.94.100501}. Still, for other higher-dimensional geometries topologically protected edge states  naturally also appear, although these states will form dispersing edge states.
This is of great practical utility when scanning for candidate topological superconductors as it requires much less stringent experimental conditions.
For example, islands of arbitrary shape and size can be studied, rather than highly specialized wire geometries.

The most direct and natural way to study real space structures, such as topological superconductor islands, is through scanning tunneling microscopy (STM) measurements.
Such an experiment was recently carried out on Pb/Co/Si(111) with clear indications of topologically protected edge states \cite{arXiv:1607.06353}. Here Pb provides superconductivity and Rashba spin-orbit interaction, while Co islands provide the necessary magnetism to seemingly generate a topological superconducting state inside the islands.
While the evidence are convincing, it is important to remember that even if the edge states are established as due to a non-trivial topology, different topological phases can result in widely differing edge states. The most prominent distinction is that between a single chiral edge state and counter-propagating, or helical, edge states. This distinction is not only crucial for establishing the basic physical properties, but more importantly, only the chiral topological superconductor can host non-degenerate Majorana bound states, providing the most direct route for constructing Majorana bound state devices.

Straightforward local density of states (LDOS) measurements cannot easily distinguish between chiral and helical edge states. However, spin-polarization around isolated magnetic impurities have recently been used as a tool to characterize different properties of spin-orbit coupled and topological superconductors \cite{PhysRevB.79.060505, PhysRevB.93.214514, PhysRevB.94.134511, PhysRevLett.115.116602}.
It has also recently been shown that a chiral topological superconductor with Rashba spin-orbit interaction exhibits persistent spin-polarized currents along its edges \cite{PhysRevB.92.214501}.
These persistent currents can be understood as a consequence of the chirality, while the spin-polarization is due to the spins in the currents coupling to the Rashba spin-orbit interaction.
Together, these results suggests that spin-polarization measurements might be very attractive for determining the nature of the edge states in candidate topological superconductors.

In this work we perform extensive numerical calculations of the LDOS and spin-polarized LDOS for ferromagnetic islands embedded in conventional $s$-wave superconductors with Rashba spin-orbit coupling within the topologically non-trivial phase.
We find that the edge states are clearly visible in the LDOS forming a characteristic x-shaped structure in real space when crossing the energy gap. This gives rise to a one-ring structure around the island at zero bias, while at higher energies the edge states naturally form a two-ring structure.
This clearly establish the existence of edge states, but it is notably not possible to determine the number of edge states from this data. 
However, our spin-polarized LDOS results clearly shows that only one spin species cross the Fermi level at any given edge, manifestly proving the chiral nature of the edge states.
In particular, we show that this branch has an in-plane spin-polarization directed along the normal to the edge.
At higher ingap energies we find that the spin-polarization is transferred to the opposite edge of the island, which provides a very powerful experimentally accessible signature for chiral edge states. 

For completeness we also investigate the influence of a $p$-wave superconducting order parameter component instead of the Rashba spin-orbit interaction, as such are often used interchangeably in many theoretical models.
However, we find no qualitative difference compared to the chiral state generated by finite spin-orbit interaction, and thus conclude that edge state properties does not change between using a $p$-wave component or a Rashba spin-orbit interaction.
In particular, this is in contrast to arguments presented in Ref.~\cite{arXiv:1607.06353}, where it was argued that a $p$-wave order parameter was needed to explain their experimental results.
All the relevant signatures are here reproduced in both models and no extension beyond the conventional spin-orbit picture is therefore needed to explain the data.

\begin{figure}
\includegraphics[width=245pt]{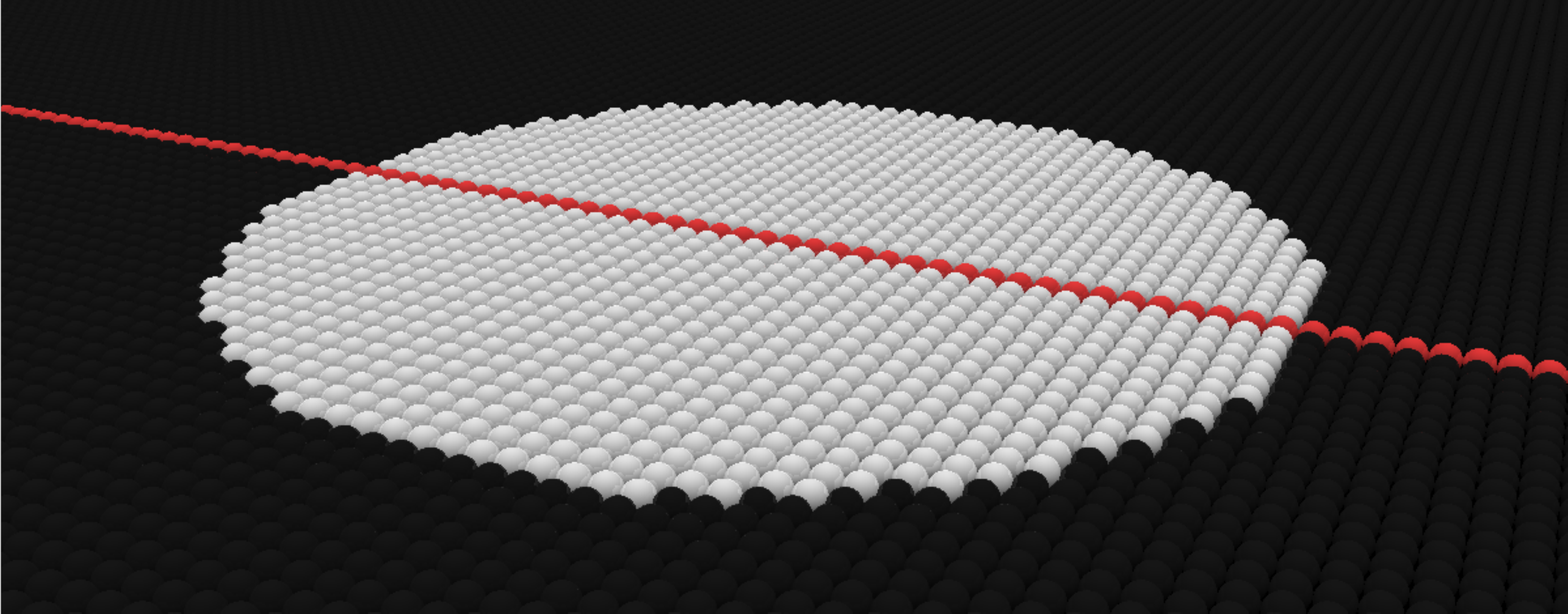}
\caption{Schematic of a circular ferromagnetic island (grey) embedded in a 2D superconductor (black) modeled using a total of $201\times 201$ sites.
The (spin-polarized) LDOS in  Fig.~\ref{Figure:LDOS} and \ref{Figure:SpinPolarizedLDOS} are calculated along the red line.
While the magnetic atoms are contained in the circular region, the effective Zeeman field is assumed to have a smooth profile across the boundary.}
\label{Figure:Schematic}
\end{figure}

We consider a general 2D superconductor with spin-orbit coupling on a square lattice with additional finite sized ferromagnet islands, see Fig.~\ref{Figure:Schematic} and described by the Hamiltonian \cite{PhysRevLett.103.020401, PhysRevLett.104.040502, PhysRevB.84.180509, PhysRevB.88.024501, PhysRevB.91.214514, PhysRevLett.115.116602, PhysRevB.92.214501, PhysRevB.94.100501}
\begin{align}
	\mathcal{H} &= \mathcal{H}_{kin} + \mathcal{H}_{\Delta_{s}} + \mathcal{H}_{so} + \mathcal{H}_{\Delta_{p}} + \mathcal{H}_{V_z}.
\label{Equation:Tight_binding_Hamiltonian}
\end{align}
The two first terms  are given by
\begin{align}
	\mathcal{H}_{kin} &= -t\sum_{\langle\mathbf{i},\mathbf{j}\rangle,\sigma}c_{\mathbf{i}\sigma}^{\dagger}c_{\mathbf{j}\sigma} - \mu\sum_{\mathbf{i},\sigma}c_{\mathbf{i}\sigma}^{\dagger}c_{\mathbf{i}\sigma}, \nonumber \\
	\mathcal{H}_{\Delta_{s}} &= \sum_{\mathbf{i}}\left(\Delta_{s} c_{\mathbf{i}\uparrow}^{\dagger}c_{\mathbf{i}\downarrow}^{\dagger} + {\rm H.c.}\right), \nonumber
\end{align}
and describe a conventional $s$-wave superconductor, where $c_{\mathbf{i}\sigma}^{\dagger}$ ($c_{\mathbf{i}\sigma}$) is a creation (annihilation) operator for a $\sigma$-spin on site $\mathbf{i}$, $t$ the nearest-neighbor hopping amplitude, $\mu$ the chemical potential, and $\Delta_{s}$ a conventional spin-singlet $s$-wave superconducting order parameter.
The next two terms are 
\begin{align}
	\mathcal{H}_{so} &= \alpha\sum_{\mathbf{i}\mathbf{b}}\left(e^{i\theta_{\mathbf{b}}}c_{\mathbf{i}+\mathbf{b}\downarrow}^{\dagger}c_{\mathbf{i}\uparrow} + {\rm H.c.}\right), \nonumber \\
	\mathcal{H}_{\Delta_{p}} &= \Delta_{p}\sum_{\mathbf{i}\mathbf{b}}\left(e^{i\theta_{\mathbf{{b}}}}c_{\mathbf{i}+\mathbf{b}\uparrow}^{\dagger}c_{\mathbf{i}\downarrow}^{\dagger} + {\rm H.c.}\right), \nonumber
\end{align}
which adds a Rashba spin-orbit interaction with strength $\alpha$ and a chiral spin-triplet $p$-wave superconducting order parameter $\Delta_{p}$, respectively. Here, $\mathbf{b}$ runs over the vectors that point along the nearest-neighbor bonds and $\theta_{\mathbf{b}}$ is its polar angle.
In a superconductor with spin-orbit interaction such an explicit $p$-wave component is present if the Cooper pairs are formed in the basis where the original kinetic energy plus the spin-orbit interaction is diagonal \cite{Alicea10}.
Finally, we get the effect of a ferromagnetic island by adding the Zeeman exchange term
\begin{align}
	\mathcal{H}_{V_z} &= -\sum_{\mathbf{i},\sigma,\sigma'}V_z(\mathbf{i})\left(\sigma_z\right)_{\sigma\sigma'}c_{\mathbf{i}\sigma}^{\dagger}c_{\mathbf{i}\sigma'}. \nonumber
\end{align}
Unlike all other terms, which are homogeneous throughout the system, the Zeeman term $V_{z}(\mathbf{i})$ is given a spatially varying value $V_z(\mathbf{i})$ that smoothly transitions from zero outside of the island, to a finite value inside the island.

We model the system described by Eq.~\eqref{Equation:Tight_binding_Hamiltonian} on a large ($201 \times 201$) square lattice with parameters (in units of $t$): $\mu = -3.9$, $\Delta_s = 0.08$, and one of $\Delta_p$ or $\alpha$ set to 0.28 while the other is set to 0.
For the ferromagnetic islands we assume the profile $V_{z}(\mathbf{i}) = 0.24(1/2 - \textrm{atan}((r-R)/W)/\pi)$, where $r$ is the distance from the center of the system, $R$ is the radius of the island, and $W = 5$ sets the scale over which the Zeeman term decays to zero at the island boundary.
For the finite spin-orbit interaction $\alpha$ model, this choice of parameters put the ferromagnetic island well within a chiral topological phase with a single chiral edge state, since the condition for the topological phase reads $(-4t+\mu)^2 + |\Delta_s|^2 < V_z^2$ \cite{PhysRevB.82.134521}. Note that without finite magnetism this model is always topologically trivial. 
In the alternative case of a finite $p$-wave pairing $\Delta_p$, a time-reversal invariant topological phase with helical edge states occurs for very large $\Delta_p$ when no magnetism is present \cite{PhysRevB.79.060505}, but we choose $\Delta_p$ such that the surrounding superconductor is decisively within the trivial phase.
We have checked that the exact value of all parameters are not of importance and our results have general qualitative validity. This is also true when varying the dimensional parameters $R$ and $W$. The particular choice of parameters have been chosen to provide quantitatively relevant results for the already experimentally realized system of Co ferromagnetic islands on Pb \cite{arXiv:1607.06353}.

To solve Eq.~\eqref{Equation:Tight_binding_Hamiltonian} we use a Chebyshev polynomial expansion method \cite{RevModPhys.78.275, PhysRevLett.105.167006, PhysRevB.94.100501} to expand the non-principal part of the Green's function $G_{\sigma\sigma'}(\mathbf{i},\mathbf{i}, E)$ using 4000 Chebyshev coefficients.
The LDOS is then calculated as
	$\rho(\mathbf{i}, E) = -\sum_{\sigma}\frac{1}{\pi}G_{\sigma\sigma}(\mathbf{i},\mathbf{i}, E)$.
Note that the imaginary part is here not taken as conventionally done when using the retarded Green's function, since the non-principal part has already been isolated in the Chebyshev expansion \cite{PhDThesisKristofer}.
Similarly, the spin-polarized LDOS along the spin-polarization axis $\mathbf{\hat{n}}$ is calculated using
	$\rho_{\mathbf{\hat{n}}}(\mathbf{i}, E) = -\frac{1}{\pi}\sum_{\sigma\sigma'}\left(\langle\mathbf{\hat{n}}|\right)_{\sigma}G_{\sigma\sigma'}(\mathbf{i},\mathbf{i}, E)\left(|\mathbf{\hat{n}}\rangle\right)_{\sigma'}$,
where
$|\mathbf{\hat{n}}\rangle = \cos(\frac{\theta}{2})|\uparrow\rangle + \sin(\frac{\theta}{2})e^{i\varphi}|\downarrow \rangle$. All calculations were implemented using the TBTK library for discrete second-quantized models \cite{TBTK, TBTK2017_09_26}.

\begin{figure}
\includegraphics[width=245pt]{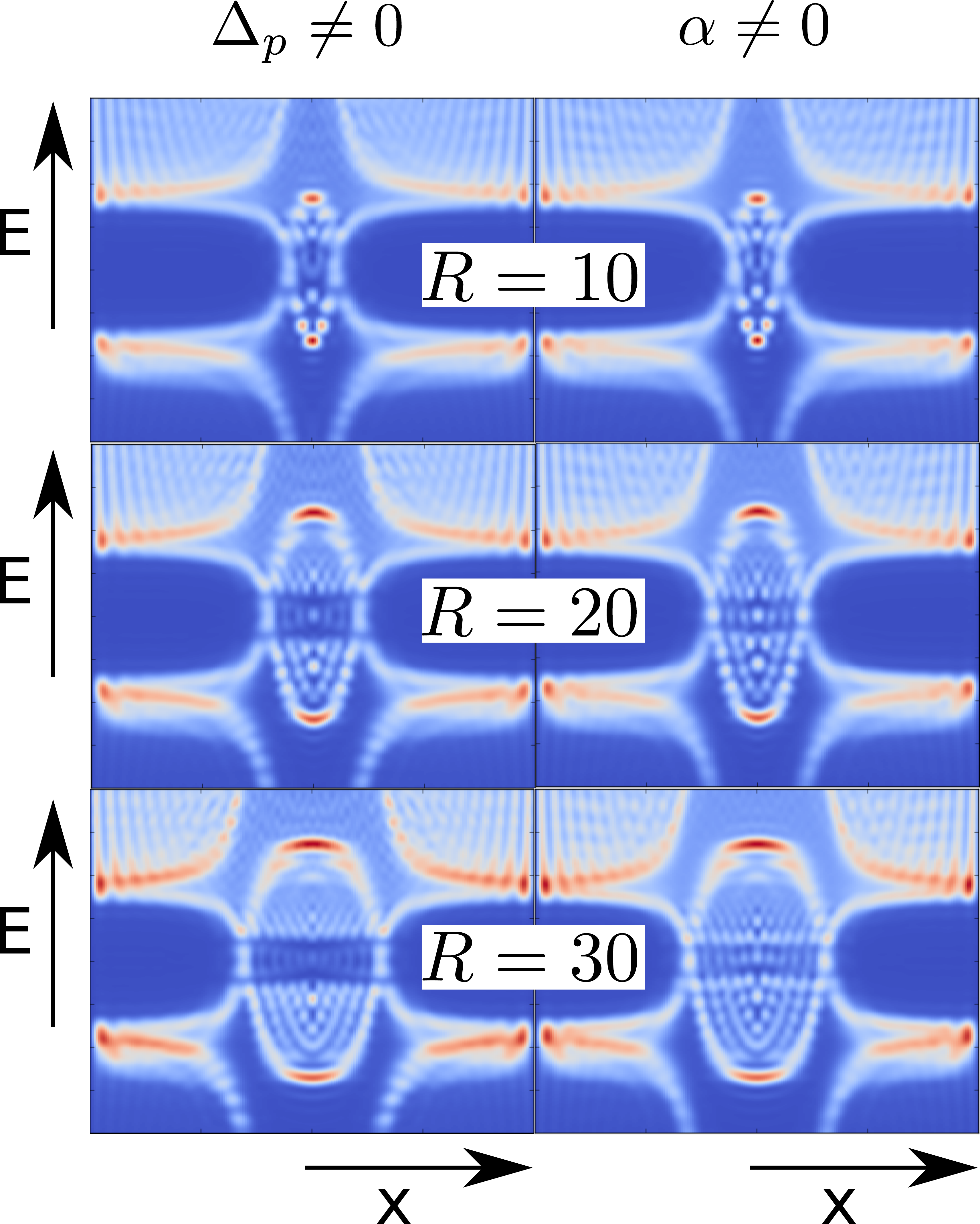}
\caption{LDOS along a line crossing through a ferromagnetic island for a system with either finite spin-triplet pairing (left) or  Rashba spin-orbit interaction (right). Blue (red) corresponds to minimum (maximum) values with color intensities normalized independently in each panel.}
\label{Figure:LDOS}
\end{figure}

Turning to the results, we plot in Fig.~\ref{Figure:LDOS} the LDOS for three different island sizes, for the case of non-zero $p$-wave superconductivity (left) or Rashba spin-orbit interaction (right).
As seen, the LDOS exhibits a clear x-shaped feature crossing through the energy gap around the edges of the ferromagnetic island, a feature that is remarkably similar for both models. 
Notably, these x-shaped states are very localized in space, which provides, even by itself, strong evidence for these states being topologically protected edge states. 
We note that this is also very similar to both the experimental and numerical results reported in Ref.~[\onlinecite{arXiv:1607.06353}].
While we know that the model with finite spin-orbit interaction have a chiral edge state, an LDOS measurement by itself can not give any information regarding whether these states are actually chiral or helical.
In fact, from the total LDOS plots in Fig.~\ref{Figure:LDOS} it is tempting to conclude that it is two branches crossing the Fermi level, which would mean that the edge states are either helical or that there are two chiral modes, both clearly incorrect for at least the system with finite spin-orbit interaction.

\begin{figure}
\includegraphics[width=245pt]{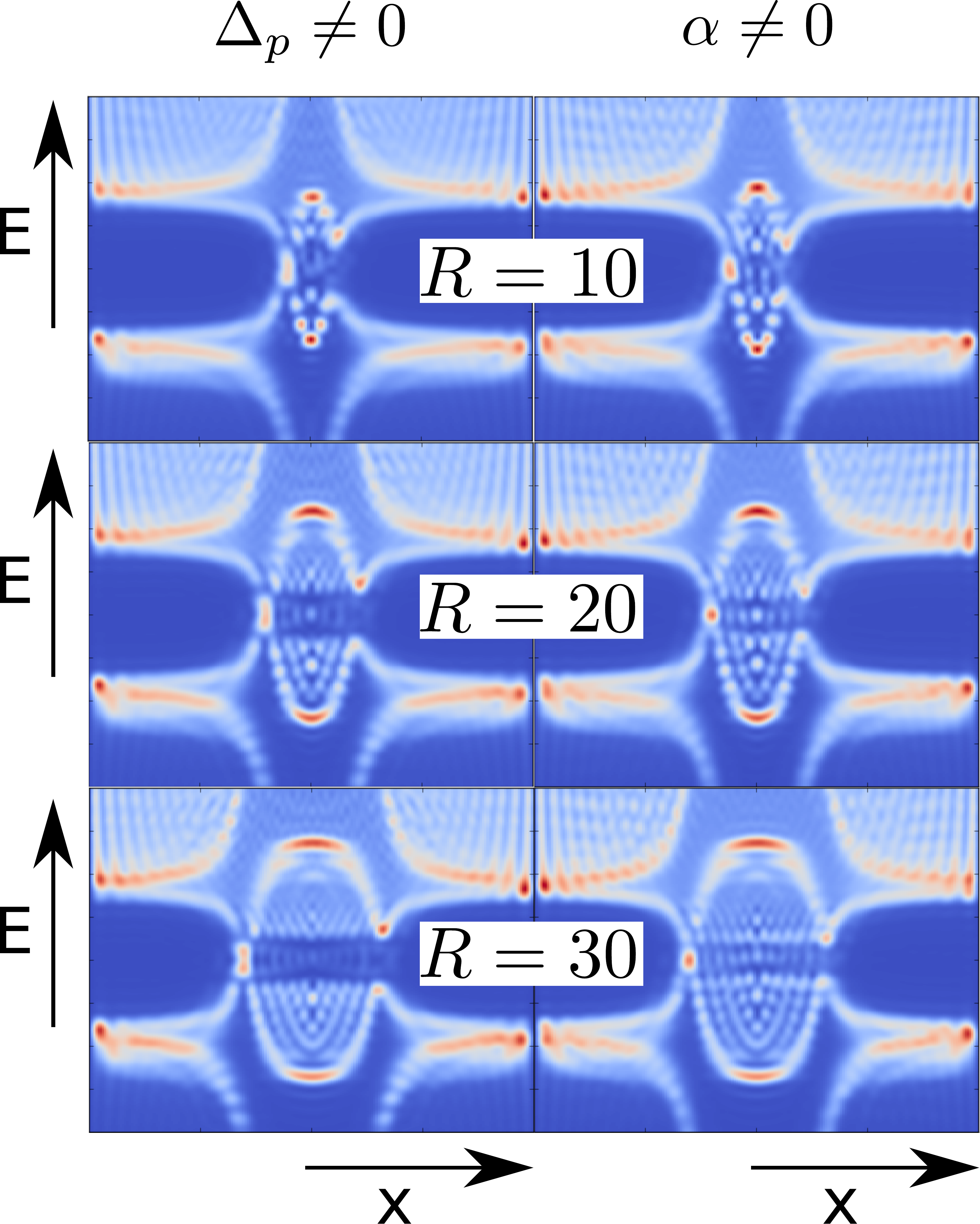}
\caption{Same as in Fig~\ref{Figure:LDOS}, but for spin-polarized LDOS with spin-polarization along the positive x-axis.}
\label{Figure:SpinPolarizedLDOS}
\end{figure}

In Fig.~\ref{Figure:SpinPolarizedLDOS} we go further and present the the spin-polarized LDOS with the spin-polarization chosen along the positive x-axis.
These plots immediately reveal that the x-up spin-polarization branch cross the Fermi surface only on one side of the sample (here left side), while this spin is completely absent on the opposite edge.
A reversal of the spin-polarization axis similarly shows that the x-down branch crosses the Fermi level on the opposite edge.
This result directly establishes that there is only one branch along the edge, which necessarily implies a single, spin-polarized, chiral edge state.

In both Figs.~\ref{Figure:LDOS}-\ref{Figure:SpinPolarizedLDOS} we see that the two models have the very same qualitative behavior for the whole range of grain sizes.
We can therefore conclude that the two models are entirely interchangeable as far as the edge state features are concerned. Notably both models produce single chiral edge states around the ferromagnetic island.
The only visible difference between the two models is that the model with Rashba spin-orbit interaction has a somewhat higher density of intragap states that pollutes the low energy spectrum inside the island.
In the following we continue with a more detailed study of the Rashba spin-orbit interaction model, but note that we have confirmed all results in both models.
We choose the spin-orbit interaction model because it has the more polluted low-energy spectrum, which means that low-energy edge state features clearly visible in this model are only clearer in the $p$-wave model.
It is also the Rashba spin-orbit interaction that is the primary source of the non-trivial topology in actual materials, since the $p$-wave superconductivity is usually induced by the Rashba spin-orbit interaction.

\begin{figure}
\includegraphics[width=245pt]{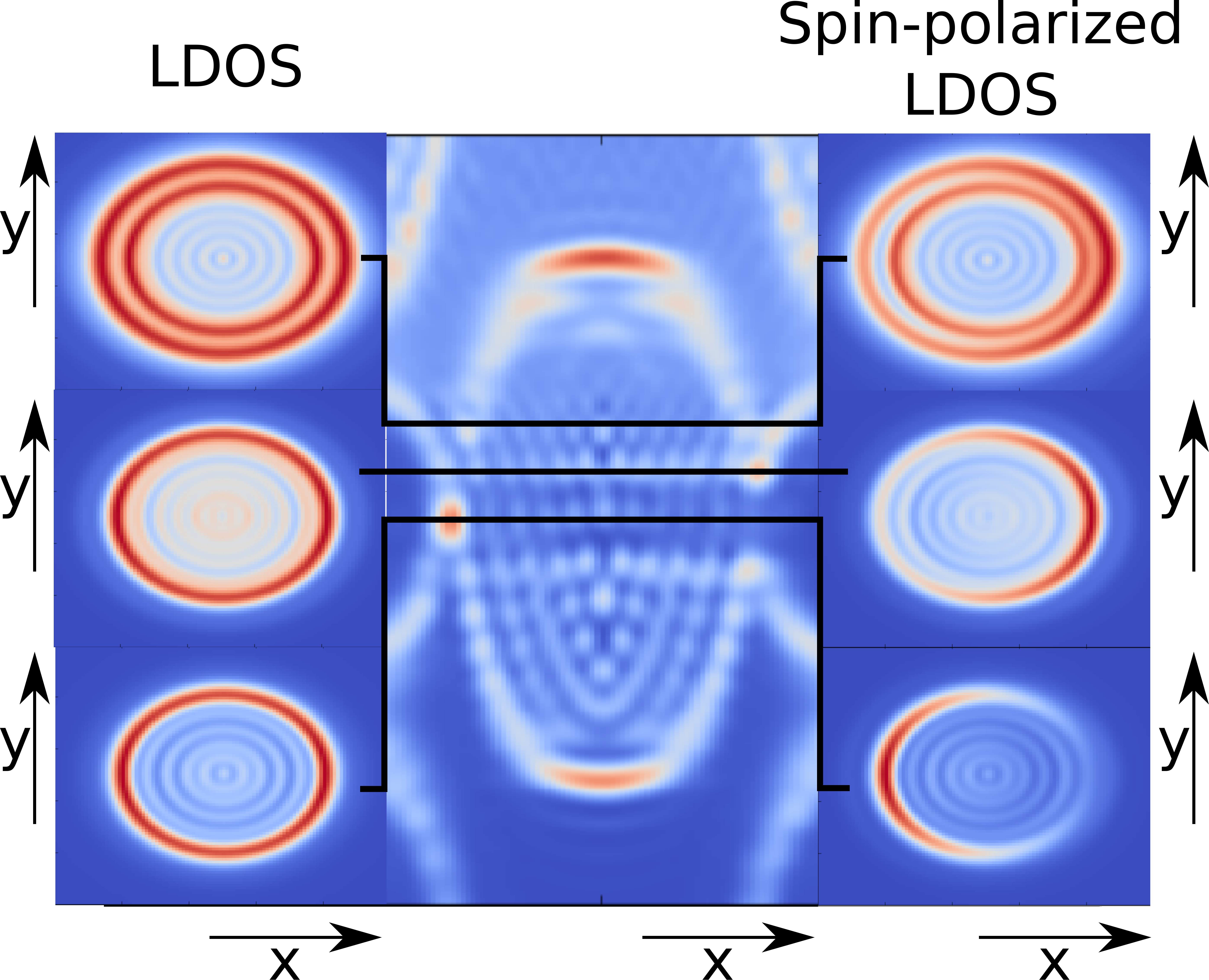}
\caption{(Bottom right panel of Fig.~\ref{Figure:SpinPolarizedLDOS} (center) with 2D topographic total LDOS (left) and spin-polarized LDOS (left) at energies indicated by the black lines: $E = 0$ (bottom), $E = 0.024$ (middle) and $E = 0.0496$ (top).}
\label{Figure:2D}
\end{figure}

To further understand the properties of the chiral edge states in terms of LDOS measurements, we plot in Fig.~\ref{Figure:2D} both the LDOS and spin-polarized LDOS over the whole surface surrounding a $R = 30$ ferromagnetic island for three different energies.
At $E = 0$ (bottom panels) the x-up spin-polarization is clearly localized on the left edge, while the total LDOS is symmetrically distributed around the whole edge.
At a higher energy (middle panels) the spin-polarized LDOS is however transfered and instead becomes mainly localized on the opposite, right edge.
This can be understod as a direct consequence of the avoided crossing for the up-spin branch at the right edge of the island (see center energy-resolved panel).  
This avoided crossing causes a locally flat dispersion and thus the spin-polarization of this gapped edge state overwhelms all other contributions starting at the avoided crossing energy. The result is a net x-up polarization on the right side of the island at this energy and thus an overall transfer of spin-polarization between the edges of the island as function of increasing energy.
We here strongly emphasize that the chiral edge state at zero energy on the right side, with its x-down spin polarization, only exists because the x-up branch have this avoided crossing and is thus fully gapped. Thus this avoided crossing should not be seen as the remnant of any helical state, but it is an intrinsic component of any chiral edge state.
A similar increased x-up spin concentration at the right edge also occur at the corresponding negative energy, although it is not displayed in the figure.

We point out that the transfer in concentration of the spin-polarized LDOS from one edge to the other as a function of the bias voltage is particularly interesting from an experimental point of view.
Namely, the presence of topologically protected chiral edge states can be detected as a characteristic oscillation of the spin-polarized LDOS from one edge to the other as the bias voltage is swept through the superconducting gap.
This is particularly useful if the edge states are much more symmetrically distributed than in Fig.~\ref{Figure:2D}, for example if $V_z$ is strong enough to almost entirely tilt the spins inside the island perpendicular to the surface.
The difference between the edges for the spin-polarized signal is then notably smaller.
The oscillating nature of the spin-polarized LDOS can be utilized as an additional signature because it provides a method for verifying whether the contrast is large enough to distinguish the values at the two different edges.
Notably this can be done without any physical modification of either system or probe, only a change in bias voltage is needed.
The oscillating spin-polarized LDOS can also be utilized to detect the chiral nature of the edge state when studying a single straight edge, rather than the edge of a circular island.

At even higher energies (Fig.~\ref{Figure:2D}, top) a two-ring structure develops symmetrically around the whole edge with the spin-polarized LDOS approaching the same symmetric appearance as the total LDOS.
Similarly, a two-ring structure also develops at the corresponding negative energies.
Such a two-ring structure has also recently been reported experimentally \cite{arXiv:1607.06353}.
We also note that, while the low-energy spectrum looks notably polluted inside the ferromagnetic island for the line cuts shown in Figs.~\ref{Figure:LDOS}-\ref{Figure:SpinPolarizedLDOS}, the full 2D plots in Figs.~\ref{Figure:2D} make it clear that the edge features at the edge are in fact clearly dominating.
Moreover, the notable ripple pattern seen in Fig.~\ref{Figure:2D} reveals that the intragap energy features are due to the tails of the edge states stretching into the island and are thus diminishing with increasing island size.

We have also performed calculations for different spin-polarization directions.
As long as the spin-polarization is in-plane, the spin-polarized LDOS shows similar results as in Fig.~\ref{Figure:2D}.
The only difference is that the position of the maximum intensity rotates together with the spin-polarization axis, such that it always occur when the edge is perpendicular to the spin-polarization axis.
However, if the spin-polarization axis is taken to be perpendicular to the surface, then the spin-polarized LDOS is practically identical to the LDOS.
This is a consequence of the spin-polarization of the edge states having its origin in the Rashba spin-orbit interaction, which has only an in-plane spin dependence.

In summary, we have shown that spin-polarized LDOS measurements can be a very powerful tool for detecting topological superconductors with chiral edge states.
For any in-plane spin-polarization axis, the spin-polarized low-energy LDOS is located on only one side of the island in the topologically non-trivial phase.
By simply sweeping the bias voltage through the gap the spin-polarization is transferred from one island side to the opposite.

\acknowledgments
We thank J.~Cayao, D.~Roditchev, and P.~Simon for useful discussions. This work was supported by the Swedish Research Council (Vetenskapsr\aa det), The Knut and Alice Wallenberg Foundation through the Wallenberg Academy Fellows program, the Swedish Foundation for Strategic Research (SSF), and the G\"{o}ran Gustafsson Foundation.

\end{document}